\documentclass[sigconf]{acmart}

\AtBeginDocument{%
  }

\setcopyright{acmlicensed}
\copyrightyear{2018}
\acmYear{2018}
\acmDOI{XXXXXXX.XXXXXXX}

\acmConference[Conference acronym 'XX]{Make sure to enter the correct
  conference title from your rights confirmation emai}{June 03--05,
  2018}{Woodstock, NY}
\acmISBN{978-1-4503-XXXX-X/18/06}

\usepackage{algorithm}
\usepackage{algorithmic}

\begin{document}

\title[The Adaptive User-centered GUI-based AutoML Toolkit]{AdaptoML-UX: An Adaptive User-centered GUI-based AutoML Toolkit for Non-AI Experts and HCI Researchers}


\author{Amr Gomaa}
\orcid{0000-0003-0955-3181}
\affiliation{%
  \institution{DFKI, Saarland Informatics Campus}
    \city{Saarbr{\"u}cken}
  \country{Germany}
}

\email{amr.gomaa@dfki.de}

\author{Michael Sargious}
\affiliation{%
  \institution{DFKI, Saarland Informatics Campus}
    \city{Saarbr{\"u}cken}
  \country{Germany}
}

\email{michael.sargious@dfki.de}

\author{Antonio Kr{\"u}ger}
\orcid{0000-0002-8055-8367}
\affiliation{%
  \institution{DFKI, Saarland Informatics Campus}
    \city{Saarbr{\"u}cken}
  \country{Germany}
}

\email{antonio.krueger@dfki.de}

\renewcommand{\shortauthors}{Gomaa et al.}

\begin{abstract}

The increasing integration of machine learning across various domains has underscored the necessity for accessible systems that non-experts can utilize effectively. To address this need, the field of automated machine learning (AutoML) has developed tools to simplify the construction and optimization of ML pipelines. However, existing AutoML solutions often lack efficiency in creating online pipelines and ease of use for Human-Computer Interaction (HCI) applications. Therefore, in this paper, we introduce AdaptoML-UX, an adaptive framework that incorporates automated feature engineering, machine learning, and incremental learning to assist non-AI experts in developing robust, user-centered ML models. Our toolkit demonstrates the capability to adapt efficiently to diverse problem domains and datasets, particularly in HCI, thereby reducing the necessity for manual experimentation and conserving time and resources. Furthermore, it supports model personalization through incremental learning, customizing models to individual user behaviors. HCI researchers can employ AdaptoML-UX (\url{https://github.com/MichaelSargious/AdaptoML_UX}) without requiring specialized expertise, as it automates the selection of algorithms, feature engineering, and hyperparameter tuning based on the unique characteristics of the data.

\end{abstract}



\begin{teaserfigure}
\centering
  \includegraphics[width=0.8\linewidth]{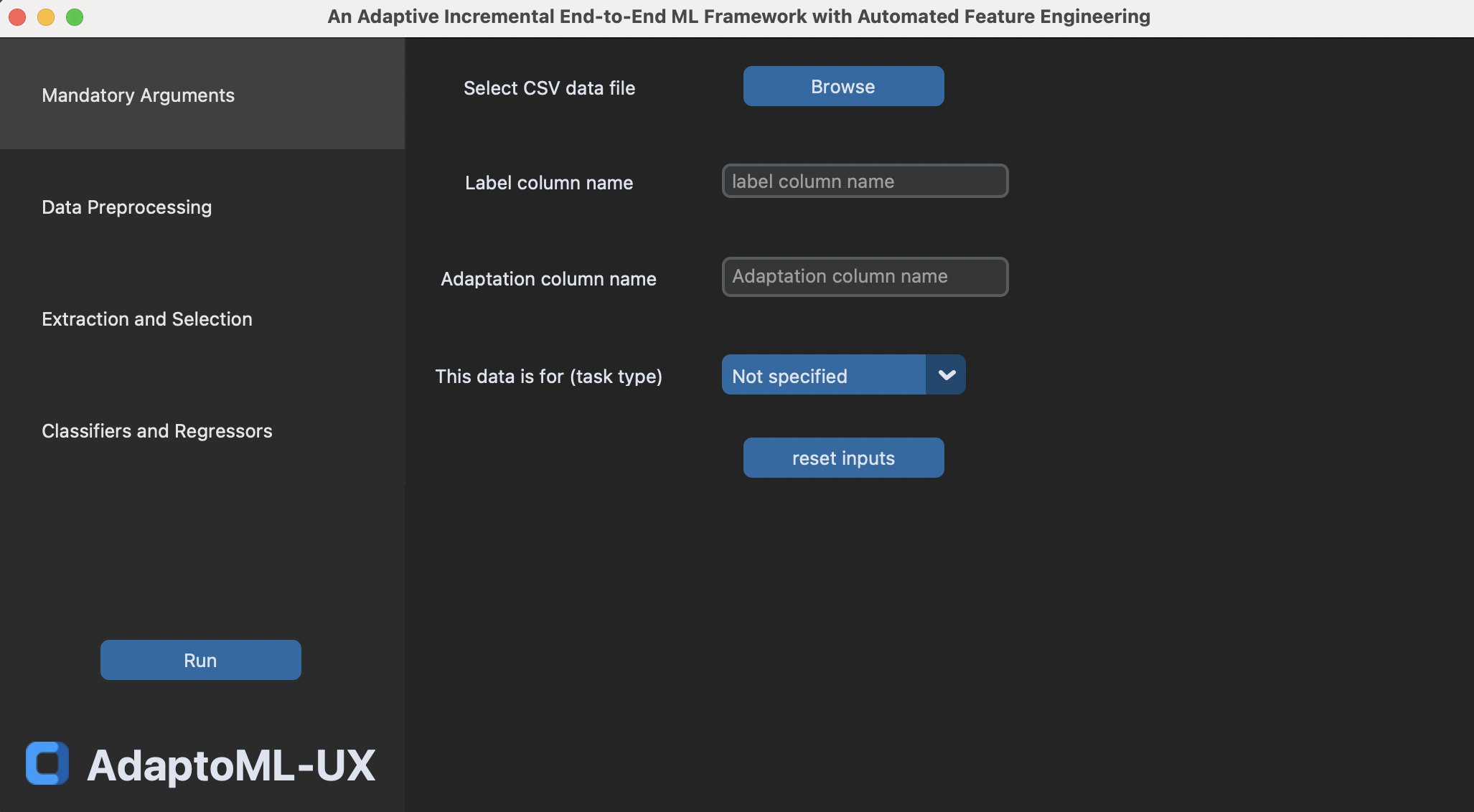}
\caption{
The \textit{AdaptoML-UX} Toolkit demonstrating the initial parameters selection, which includes four compulsory arguments: the dataset file, the labels column, the adaptation (e.g., a user-centered personalization) column, and whether a regression or classification task is required. If the remaining varaibles are not specified, the tool will conduct a grid search over all possible machine learning models and hyperparameters automatically.
}
  \Description{We present, a new adaptive automated machine learning toolkit, \textit{AdaptoML-UX}, that automates critical steps, including algorithm selection, feature engineering, and hyperparameter tuning. The depicted image shows a snapshot of the first few selected parameters in the toolkit.}
  \label{fig:teaser}
\end{teaserfigure}

\maketitle

\section{Introduction}

In today’s data-driven world, machine learning models are extensively used across various domains. While several automated machine learning (AutoML) tools exist~\cite{feurer2015efficient,thornton2013auto,komer2014hyperopt,olson2016tpot}, they are primarily designed for experts in machine learning (ML) and software engineering~\cite{karmaker2021automl}. However, the interdisciplinary nature of human-computer interaction (HCI) research, which involves professionals from diverse fields such as psychology, journalism, and medicine, necessitates a more accessible toolkit for non-AI experts. 
To address this need, we introduce \textit{AdaptoML-UX}, a toolkit specifically designed for non-specialists. It empowers users to control the machine learning pipeline and understand the underlying steps and modules without requiring extensive artificial intelligence (AI) or programming knowledge. Unlike existing AutoML frameworks, our user-friendly graphical design eliminates complex programming and debugging steps while providing clear visibility into the pipeline implementation (see~\autoref{fig:teaser}). This design ensures that users from various backgrounds can effectively engage with and utilize machine learning techniques.
Furthermore, our toolkit incorporates incremental learning and model adaptation approaches, utilizing user-centered adaptive machine learning techniques specifically designed for human-centered artificial intelligence (HAI)~\cite{gomaa2023towards,gomaa2021ml,vesin2018learning,gomaa2022adaptive}. These features enable the toolkit to adapt to changing data distributions and user-specific behaviors, ensuring that the models remain relevant and accurate over time.
In conclusion, \textit{AdaptoML-UX}~\footnote{\url{https://github.com/MichaelSargious/AdaptoML_UX}} automates crucial steps such as algorithm selection, feature engineering, and hyperparameter tuning. Additionally, it incorporates adaptive capabilities and continual learning techniques, allowing for the development of personalized models that can adapt to dynamic data distributions and user-specific behavior. By making advanced machine learning techniques accessible to non-specialists, our toolkit bridges the gap between AI technology and interdisciplinary HCI research, fostering innovation and collaboration across diverse fields.

\section{Related Work}

Our toolkit addresses several issues in current AutoML frameworks, including ease of use, a graphical user interface (GUI) for non-programmers, and automatic model adaptation and continual learning. We demonstrate these aspects in the following sections: ``Existing AutoML Frameworks'' and ``Existing Incremental Learning and Model Adaptation Toolkits''.

\subsection{Existing AutoML Frameworks}

In recent years, significant progress has been made in the field of AutoML, with various frameworks developed to streamline machine learning model selection, hyperparameter tuning, and architectural exploration~\cite{thornton2013auto,feurer2015efficient,komer2014hyperopt,olson2016tpot,celik2022online}. These frameworks aim to simplify machine learning for novices, reduce the time and resources needed to build high-performing models, and accelerate the adoption of machine learning across different fields. Thornton et al.~\cite{thornton2013auto} introduced Auto-WEKA, which focuses on automating the selection of learning algorithms and hyperparameter configurations using Bayesian optimization. Similarly, Feurer et al.~\cite{feurer2015efficient} developed AUTO-SKLEARN, an efficient AutoML framework built on scikit-learn. It includes numerous classifiers and preprocessing techniques, leveraging past performance data to improve model selection. Komer et al.~\cite{komer2014hyperopt} presented Hyperopt-sklearn, which uses the Hyperopt library to optimize algorithm configurations within the Scikit-learn library. Alternatively, Olson et al.~\cite{olson2016tpot} introduced TPOT, an AutoML framework that uses genetic programming to optimize machine learning pipelines. TPOT aims to enhance classification accuracy by evolving pipelines, but the genetic programming approach can be difficult for non-specialists to grasp. Finally, Celik et al.~\cite{celik2022online} created OAML to streamline pipeline construction for online learning and handle data drift. It explores various configurations of online learners and preprocessing methods. 

While these frameworks have proven valuable in various academic disciplines, they often pose challenges for non-expert users due to their complexity, extensive configuration options, and scripting-based methodologies that can overwhelm users without a strong ML and programming background. Additionally, unlike AdaptoML-UX, these methods are not capable of performing model adaptation and user personalization. AdaptoML-UX distinguishes itself by offering exceptional user-friendliness, specifically designed for non-experts in Human-Computer Interaction (HCI). It provides a graphical user interface (GUI) and automates model adaptation and continual learning, making advanced machine learning accessible to a broader audience.

\subsection{Existing Incremental Learning and Model Adaptation Toolkits}

Incremental learning allows machine learning models to continuously update and improve as new data becomes available, without discarding previously learned knowledge. This capability is crucial for real-world applications where data is constantly evolving. Montiel et al.~\cite{montiel2018scikit} developed Scikit-Multiflow to facilitate knowledge acquisition from data streams and multi-output learning. It integrates advanced learning techniques, data generators, and evaluators for diverse stream-learning scenarios. Despite its robust capabilities, the framework is highly modularized and requires extensive programming to use. Later, the same researchers created River~\cite{montiel2021river}, a framework designed to handle dynamic data streams and facilitate continual learning. River combines insights from the Creme~\cite{creme} and Scikit-Multiflow approaches to create a comprehensive framework for machine learning on continuously generated data. However, it suffers from the same drawbacks as previous approaches in terms of ease of use and simplicity in design.

HCI researchers have also studied several design approaches for continual learning and personalization using their own tools. Putze et al.~\cite{putze2021multimodal} focused on identifying and mitigating obstacles in human-computer interactions (HCI) using various data modalities. They emphasized the importance of customization and adaptation methodologies in HCI, suggesting that understanding human cognitive processes can lead to better adaptation techniques. Rudovic et al.~\cite{rudovic2018personalized} explored the use of robots to aid children with autism during therapy sessions, highlighting the potential for personalized machine learning frameworks tailored to individual needs. Finally, Gomaa et al.~\cite{gomaa2021ml} focused on multimodal interaction for referencing objects outside a vehicle, using gaze and pointing gestures. They demonstrated the importance of considering individual behavioral variations to design adaptable and personalized systems. However, these approaches often pose challenges in dynamically changing environments due to their complexity and the need for extensive customization.
In contrast, our \textit{AdaptoML-UX} toolkit utilizes all of the previously mentioned features, including incremental learning techniques, model adaptation, and personalization, while being designed to be as generic as possible. This allows it to be used across multiple domains and different data sources through a user-friendly interface.

\begin{algorithm}
\caption{AdaptoML-UX Pseudo Code}
\label{alg:one}

\begin{flushleft}
\textbf{Input}: \newline 
$\mathbf{D}$: Path of the data file \newline 
$\mathbf{Per}_{col\_name}$: Name of personalization column \newline 
$\mathbf{X}_{pred}$: Path of data used for prediction \newline 
$\mathbf{X}_{inc}$: Path of data used for partial fit \newline 
$model\_paths$: List of paths potining to already trained models \newline

\textbf{Output}: Results and evaluation reports represented in CSV tables and graphs \newline
\end{flushleft}

\begin{algorithmic}[1]

\STATE \textbf{Initialize:} Load dataset $\mathbf{D}$ and configuration parameters

\IF {$imputation\_flag$}
    \STATE $\mathbf{D}_{imp} \leftarrow impute(\mathbf{D})$
\ELSE
    \STATE $\mathbf{D}_{imp} \leftarrow \mathbf{D}$
\ENDIF

\STATE $\mathbf{X}, \mathbf{Y} \gets splitting\_data(\mathbf{D}_{imp})$

\IF {$feature\_extraction$}
    \STATE $\mathbf{f}_{processed} \leftarrow feature\_extract(\mathbf{X}, \mathbf{Y})$
\ELSIF {$feature\_selection$}
    \STATE $\mathbf{f}_{processed} \leftarrow feature\_select(\mathbf{X}, \mathbf{Y})$
\ELSE
    \STATE $\mathbf{f}_{processed} \leftarrow \mathbf{X}$
\ENDIF

\IF {$classifiers$ \OR $regressors$}
    \STATE create $model$ instance of classifiers or regressors
    \IF {$model\_adaptation$}
        \STATE $model.adapting\_models(\mathbf{f}_{processed}, \mathbf{Per}_{col\_name})$
    \ENDIF
    \IF {$trained\_models$}
        \STATE $model.loading\_models(model\_paths)$
    \ENDIF
    \IF {$fit$}
        \STATE $model.fit(\mathbf{f}_{processed})$
    \ENDIF
    \IF {$predict$}
        \STATE $\mathbf{Y}_{pred} \gets model.predict(\mathbf{X}_{pred})$
    \ENDIF
    \IF {$partial\_fit$}
        \STATE $model.partial\_fit(\mathbf{X}_{inc})$
    \ENDIF
\ENDIF

\end{algorithmic}
\end{algorithm}

\section{Method}

In this section, we outline the methodology employed in the development and design of the AdaptoML-UX toolkit. The overall workflow of the toolkit involves several key steps, including data imputation, feature selection, feature engineering, model selection, hyperparameter tuning, incremental learning, and automated model adaptation for various participants in user studies. 
AdaptoML-UX is implemented in Python, incorporating essential machine learning libraries to provide a robust framework for adaptive machine learning.
The design of AdaptoML-UX focuses on a user-friendly interface that guides users through the machine learning process. The user need to identify four mandatory parameters: the dataset file, the labels column, the adaptation (e.g., a user-centered personalization) column, and whether a regression or classification task is required (see~\autoref{fig:teaser}).

\begin{figure*}[t]
    \centering
    \includegraphics[width=\linewidth]{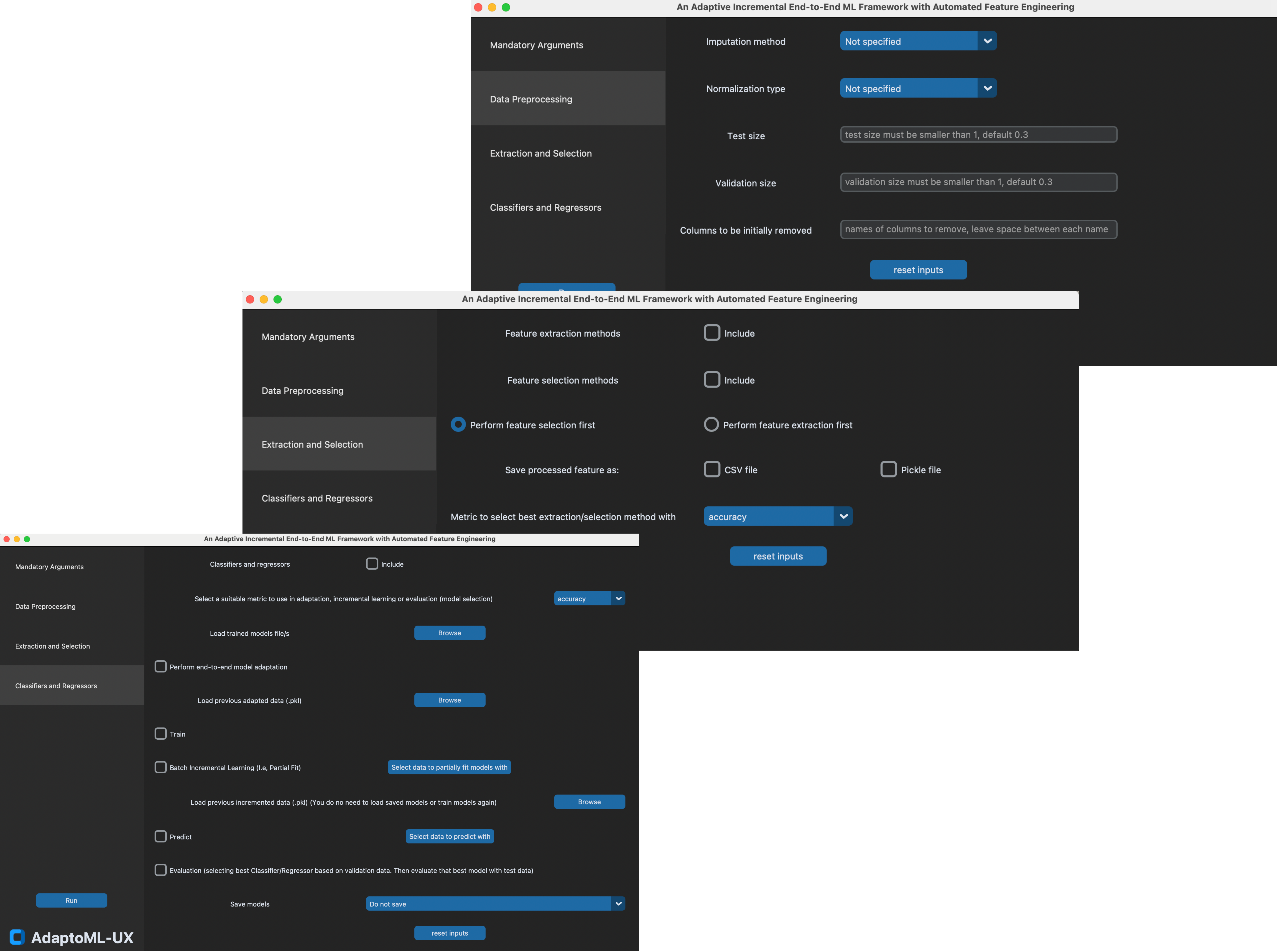}
    \caption{Snapshots of the different input parameters supported by the \textit{AdaptoML-UX} interface.}
    \Description{The image contains three snapshots from the \textit{AdaptoML-UX} interface highlighting the different selectable parameters such as whether to perform train only or prediction and train, and whether to perform feature extraction, or selection or both, and so on.}
    \label{fig:example_input_interface}
\end{figure*}

AdaptoML-UX begins with the imputation step, addressing any missing values in the data. Following this, it performs feature selection or extraction based on user-defined flags, ensuring that the most relevant features are utilized for model training. It automatically perform a grid search over various preprocessing methods, machine learning models, and  hyperparameter similar to previous AutoML tools. Then, unlike previous AutoML frameworks, the toolkit can automatically perform an extra step of model adaptation and personalization based on certain users or user groups using incremental learning approaches. The GUI is designed to simplify the workflow, allowing users to easily configure and execute these steps without requiring specialized knowledge in machine learning. Additionally, the AdaptoML-UX framework offers users a range of configurable parameters for data preproceesing, feature engineering, and supervised learning methods. In the ``Data Preprocessing Section'', users can set data imputation methods, define test and validation data sizes, and apply normalization procedures. The ``Feature Engineering Section'' allows users to enable feature extraction and selection, and store processed features in \textit{CSV} or \textit{Pickle} formats. The ``Classifiers and Regressors Section'' provides options for selecting classification or regression tasks, loading pre-trained models, activating model adaptation, specifying model fitting parameters, and utilizing incremental learning. Users can also choose the most suitable classifier or regressor based on assessment criteria and specify data for prediction and partial model fitting.~\autoref{fig:example_input_interface} highlights an overview of these different sections in the AdaptoML-UX interface automating the end-to-end machine learning pipeline and hyperparameters tuning as backend processes.

\begin{figure*}[t]
    \centering
    \includegraphics[width=\linewidth]{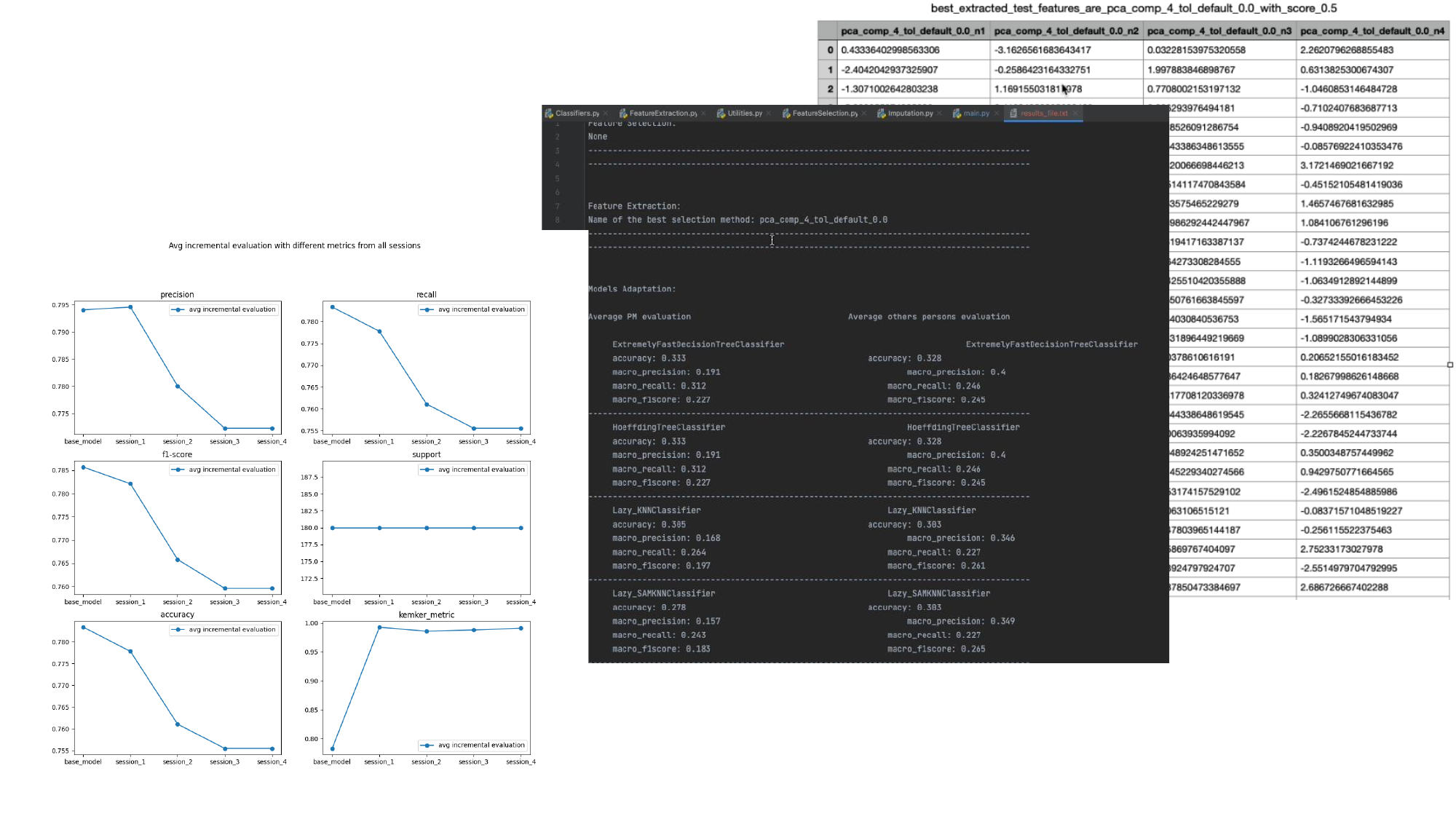}
    \caption{An example of a sample of output files (e.g., tabular results, output text, and metrics' graphs) with different metric results after training batch incremental learning over four sessions using the dataset from Gomaa et al.~\cite{gomaa2021ml}.}
    \Description{The image contains six line graphs arranged in a 3x2 grid, each representing the performance of an incremental evaluation model across different evaluation metrics and sessions over distinct time intervals. The metrics are precision, recall, F1-score, support, accuracy, and kemker loss. Each graph displays the performance of the model over five sessions (a base model and additional four sessions coming thereafter). The lines fluctuate between sessions, showing how the model’s performance changes over time for each metric. It also shows a snapshot of tabular data and text output.}
    \label{fig:example_output_results}
\end{figure*}

Throughout the process, AdaptoML-UX ensures continuous adaptation and monitoring of model performance. Users can leverage this adaptive capability to handle dynamic data distributions and maintain model accuracy over time. The pseudo code in algorithm \ref{alg:one} outlines the framework's structure, emphasizing its step-by-step approach to making machine learning accessible and efficient for non-experts. Finally, our toolkit automatically saves the machine learning models, the tabular results, and plots performance metrics graphs to evaluate the output as seen in~\autoref{fig:example_output_results}.

\section{Results}

To evaluate our toolkit, we recruited researchers with diverse backgrounds and varying levels of machine learning experience to assess the usability design goals of AdaptoML-UX. The participants provided valuable context for evaluating the interface. For instance, P1 had a foundational background in computer science with no knowledge of machine learning, while P2 specialized in Natural Language Processing (NLP) and was experienced in ML but faced challenges with AutoML. P3 focused on bioinformatics and had substantial experience in developing automated ML tools. P4 was an HCI researcher with experience in user interface design but no ML experience, and P5 was an HCI researcher knowledgeable in both ML and user interface design. This diverse expertise provided a comprehensive evaluation of AdaptoML-UX's usability across different levels of machine learning experience and various domains.

As for the evaluation metrics, we used the usability goals designed by Preece et al.~\cite{preece1994human}, including ease of learning, ease of remembering, efficiency, effectiveness, and utility. The user study aimed to determine whether AdaptoML-UX aligned with these goals. Efficiency was assessed by how quickly and accurately users could complete tasks, while effectiveness focused on the framework's ability to deliver reliable results during model updates and data incorporation. Utility was evaluated based on the overall value and practicality of the toolkit in meeting user needs. Ease of learning examined how quickly participants could understand and use the framework, and ease of remembering looked at how well users could recall and apply their knowledge after initial interactions. These findings provided valuable insights into the usability and performance of AdaptoML-UX, guiding its further development and potential integration into the field of machine learning.
To assess these usability goals, we designed three tasks. The first task involved configuring a model to adapt to evolving features, evaluating the tool's adaptability and efficiency. The second task required updating an existing model with new data without retraining, assessing ease of use and model accuracy maintenance. The final task focused on automating feature engineering, testing the tool's efficiency and utility in data preprocessing. These tasks provided practical scenarios to evaluate how well AdaptoML-UX met its design goals, offering insights for further improvements.

The qualitative feedback from participants provided valuable insights into the usability and potential improvements for AdaptoML-UX. P1 found the tool user-friendly but highlighted the need to better understand the connections between various configuration options. We addressed this by adding hints when hovering over different parameters. P2 raised concerns about the lack of clarity on default parameter values and suggested a more organized interface, including explicit indicators for loaded models and better placement of the "Run" icon. These changes have been implemented in the current version. P3 emphasized the need for transparency in the model adaptation option and enabling the training of only the highest-performing model. Although this is a backend change for training automation and does not affect the GUI, it has been implemented. P4 described the process as relatively uncomplicated but hinted at possible slight improvements regarding some of the button positions. P5 reported a smooth interaction with no significant obstacles, indicating a generally favorable experience with the framework.

Additionally, quantitative feedback was obtained through a 5-point Likert scale highlighting the usability of AdaptoML-UX across several goals. Participants rated the framework's effectiveness positively, with scores ranging from three to five. P1 and P2 rated it four, appreciating the control and transparency, while P3 gave it three, noting some uncertainty about parameter impacts. P4 rated it a perfect five for its effectiveness. Efficiency received high praise, with P1, P2, and P3 giving perfect scores of five, commending the automatic data visualization and streamlined interface. P4 and P5 also rated it highly, with scores of four, acknowledging its minimized effort and speedy usability. The utility was highly favored, with Participants 2, 3, and 4 awarding perfect fives, appreciating the comprehensive toolkit for various ML tasks. P1 and P5 rated it four, recognizing its valuable functionalities but suggesting further exploration. For ease of use, P3 and P5 gave perfect scores of five, while P4 rated it four, and P1 and P2 rated it three, pointing out areas for UI improvement. All participants unanimously rated the framework a perfect five for ease of remembering, citing its simplicity, limited options, and practicality.

\section{Conclusion and Limitations}

\textit{AdaptoML-UX} is as a user-friendly toolkit designed to bridge the gap between advanced machine learning techniques and non-expert users. Its graphical user interface (GUI) simplifies the machine learning process, making it accessible to individuals without extensive programming knowledge. \textit{AdaptoML-UX} features, such as automatic data visualization and streamlined interface, have been well-received by participants. The qualitative feedback highlights the framework's ease of use, transparency, and efficiency, with suggestions for minor improvements already being addressed. However, despite these encouraging results, the study's findings are limited by the small sample size of participants, which may affect the generalizability of the results. To gain a more comprehensive understanding of AdaptoML-UX's usability and effectiveness, further assessment involving a larger and more diverse sample from the human-computer interaction (HCI) community is necessary. Such broader evaluations will help validate the initial positive feedback and identify any additional areas for improvement, ensuring better user experience for future use.


\bibliographystyle{ACM-Reference-Format}
\bibliography{sample-base}


\end{document}